\pdfoutput=1 

\documentclass[11pt]{article}
\usepackage{amssymb}

\usepackage{amsthm} 

\usepackage{tikz}
\usetikzlibrary{trees}
\usepackage{qtree}

\usetikzlibrary{decorations.pathmorphing}
\usetikzlibrary{decorations.markings}
\usepackage{verbatim}

\usepackage{amsmath}
\usepackage{amsthm} 
\usepackage{amssymb,amsfonts,textcomp}
\usepackage{array}
\usepackage{hhline}
\usepackage{graphicx}

\usepackage[T3,T1]{fontenc} 
\usepackage{txfonts} 

\usepackage{latexsym} 
\usepackage{mathrsfs} 

\usepackage{MnSymbol} 

\usepackage{tikz}
\usetikzlibrary{arrows,shapes,chains,matrix,positioning,scopes}
\usetikzlibrary{decorations.pathmorphing}
\usetikzlibrary{decorations.markings}
\usepgflibrary{plotmarks}
\usetikzlibrary{patterns}

\usepackage{pdfpages}

\usetikzlibrary{mindmap}

\usepackage{graphicx}

\usepackage{multirow}
\usepackage{multicol}

\usepackage{subfig} 

\usepackage{float}
\usepackage{wrapfig}
\restylefloat{figure}
\restylefloat{table}
\usepackage{color, colortbl} 

\usepackage{float}
\restylefloat{figure}

\theoremstyle{definition}

\definecolor{currentcolor}{rgb}{0.8 0.4 0.2}

\tikzstyle{stochasticjumpstyle}=[diamond,draw,fill=white,>=latex,>->,dashed]
\tikzstyle{stochasticPathstyle}=[>=latex,>->,dashed]
\tikzstyle{stochasticNodestyle}=[ellipse,inner sep=1pt,text=.,fill=.!20]
\tikzstyle{blankstyle}=[fill=white,inner sep=1pt]

\def\SnakeSegLen{0.6em}
\def\SnakeAmp{0.11em}
\def\PrePostLen{5mm}

\tikzstyle{sendstyle}=[dashed,line width=1.1pt]

\tikzstyle{splitstyle}=[circle,draw]

\tikzstyle{receivestyle}=[>->,line width=1.1pt,decorate, decoration={zigzag,segment length=\SnakeSegLen, amplitude=\SnakeAmp, pre length=\PrePostLen, post=curveto, post length=\PrePostLen},text=black]

\tikzstyle{receivesigstyle}=[draw,inner sep=2pt,fill=pink!20]
\tikzstyle{receivesigstyle3}=[draw,inner sep=2pt, fill=white]
\tikzstyle{receivesigstyle2}=[ellipse,shade, draw,double,fill=red!10]

\tikzstyle{sendsigstyle}=[diamond,draw,inner sep=1pt, text=black, fill=yellow!80]
\tikzstyle{sendsigstyle3}=[circle,draw, ball color=white]
\tikzstyle{sendsigstyle2}=[diamond,draw,double, inner sep=1pt, fill=white]

\tikzstyle{snakesendstyle}=[*->, decorate, decoration={snake, segment length=\SnakeSegLen, amplitude=\SnakeAmp,  pre length=\PrePostLen, post=curveto, post length=\PrePostLen}]

\tikzstyle{snakesendstyle1}=[line width=1.1pt, decorate, decoration={snake,segment length=\SnakeSegLen, amplitude=\SnakeAmp}]

\tikzstyle{snakesendstyle3}=[decorate, decoration={markings, mark=at position .75 with {\arrow[red,line width=5mm]{>}}, snake, segment length=\SnakeSegLen, amplitude=\SnakeAmp,  pre length=\PrePostLen, post=curveto, post length=\PrePostLen}]

\tikzstyle{snakesendstyle2}=[decorate, decoration={ zigzag,segment length=\SnakeSegLen, amplitude=\SnakeAmp, line around/.style={decoration={pre length=\PrePostLen,post length=\PrePostLen}}}]

\newcounter{foo}
\makeatletter
\AtBeginDocument{\let\nameref\HyPsd@nameref}

\makeatother

\colorlet{anglecolor}{green!50!black}

\definecolor{darkgreen}{rgb}{0 0.6  0}

\definecolor{turquoise}{rgb}{0 0.41 0.41}
\definecolor{rouge}{rgb}{0.79 0.0 0.1}
\definecolor{vert}{rgb}{0.15 0.4 0.1}
\definecolor{mauve}{rgb}{0.6 0.4 0.8}
\definecolor{violet}{rgb}{0.58 0. 0.41}
\definecolor{orange}{rgb}{0.8 0.4 0.2}
\definecolor{bleu}{rgb}{0.39, 0.58, 0.93}

\definecolor{darkross}{rgb}{0.008,0.412,0.471}
\definecolor{middleross}{rgb}{0.012,0.580,0.663}
\definecolor{lightross}{rgb}{0.016,0.749,0.855}
\definecolor{darkblue}{rgb}{0.067,0.008,0.471}
\definecolor{middleblue}{rgb}{0.094,0.012,0.663}
\definecolor{lightblue}{rgb}{0.122,0.016,0.855}
\definecolor{darkpurple}{rgb}{0.471,0.008,0.412}
\definecolor{middlepurple}{rgb}{0.663,0.012,0.580}
\definecolor{lightpurple}{rgb}{0.855,0.016,0.749}
\definecolor{darkbrown}{rgb}{0.471,0.067,0.008}
\definecolor{middlebrown}{rgb}{0.663,0.094,0.012}
\definecolor{lightbrown}{rgb}{0.855,0.122,0.016}
\definecolor{darkolive}{rgb}{0.412,0.471,0.008}
\definecolor{middleolive}{rgb}{0.580,0.663,0.012}
\definecolor{lightolive}{rgb}{0.749,0.855,0.016}
\definecolor{darkgreen}{rgb}{0.008,0.417,0.067}
\definecolor{middlegreen}{rgb}{0.012,0.663,0.094}
\definecolor{lightgreen}{rgb}{0.016,0.855,0.122}
\definecolor{darkocre}{rgb}{0.471,0.298,0.008}
\definecolor{middleocre}{rgb}{0.663,0.420,0.012}
\definecolor{lightocre}{rgb}{0.855,0.541,0.016}

    \definecolor{lightblue}{rgb}{0,0,.7}
    \definecolor{orange}{rgb}{1,.7,0}
    \definecolor{darkorange}{rgb}{1,.4,0}
    \definecolor{darkgreen}{rgb}{0,.5,0}
    \definecolor{darkblue}{rgb}{0,0,.4}
    \definecolor{darkred}{rgb}{.4,0,0}
    \definecolor{gray}{rgb}{.2,.2,.2}
    \definecolor{darkgray}{rgb}{.2,.2,.2}
    \definecolor{shadecolor}{gray}{0.925}
    
\definecolor{darkred}{rgb}{0.65,0,0}
\definecolor{darkblue}{rgb}{0,0,.65}
\definecolor{darkgreen}{rgb}{0,0.5,0}
\definecolor{orange}{rgb}{1,.75,.25}
\definecolor{aqua}{rgb}{0,.25,.75}
\definecolor{grey}{rgb}{.5,.5,.5}
\definecolor{brown}{rgb}{.51,.35,.18}

\definecolor{lightblue}{rgb}{.3,.5,1}
\definecolor{orange}{rgb}{1,.7,0}
\definecolor{darkorange}{rgb}{1,.4,0}
\definecolor{darkgreen}{rgb}{0,.4,0}
\definecolor{darkblue}{rgb}{0,0,.4}
\definecolor{darkred}{rgb}{.56,0,0}
\definecolor{gray}{rgb}{.3,.3,.3}
\definecolor{darkgray}{rgb}{.2,.2,.2}

\definecolor{blue}{rgb}{0,0,1}
\definecolor{red}{rgb}{1,0,0}
\definecolor{pink}{rgb}{.933,0,.933}
\definecolor{green}{rgb}{0.133,0.545,0.133}
\definecolor{shadecolor}{gray}{0.925}

\definecolor{DarkBlue}{rgb}{0.000,0.000,0.545}
\definecolor{DarkChocolate}{rgb}{0.400,0.200,0.000}
\definecolor{DarkCyan}{rgb}{0.000,0.545,0.545}
\definecolor{DarkGoldenrod}{rgb}{0.720,0.525,0.044}
\definecolor{DarkGray}{rgb}{0.664,0.664,0.664}
\definecolor{DarkGreen}{rgb}{0.000,0.392,0.000}
\definecolor{DarkGrey}{rgb}{0.664,0.664,0.664}
\definecolor{DarkKhaki}{rgb}{0.740,0.716,0.420}
\definecolor{DarkLavender}{rgb}{0.400,0.200,0.600}
\definecolor{DarkMagenta}{rgb}{0.545,0.000,0.545}
\definecolor{DarkOliveGreen}{rgb}{0.332,0.420,0.185}
\definecolor{DarkOrange}{rgb}{1.000,0.550,0.000}
\definecolor{DarkOrchid}{rgb}{0.600,0.196,0.800}
\definecolor{DarkPeriwinkle}{rgb}{0.400,0.400,1.000}
\definecolor{DarkPurpleBlue}{rgb}{0.400,0.000,0.800}
\definecolor{DarkRed}{rgb}{0.545,0.000,0.000}
\definecolor{DarkRoyalBlue}{rgb}{0.000,0.200,0.800}
\definecolor{DarkSalmon}{rgb}{0.912,0.590,0.480}
\definecolor{DarkSeaGreen}{rgb}{0.560,0.736,0.560}
\definecolor{DarkSlateBlue}{rgb}{0.284,0.240,0.545}
\definecolor{DarkSlateGray}{rgb}{0.185,0.310,0.310}
\definecolor{DarkSlateGrey}{rgb}{0.185,0.310,0.310}
\definecolor{DarkSmoke}{rgb}{0.920,0.920,0.920}
\definecolor{DarkTurquoise}{rgb}{0.000,0.808,0.820}
\definecolor{DarkViolet}{rgb}{0.580,0.000,0.828}
\definecolor{DeepPink}{rgb}{1.000,0.080,0.576}
\definecolor{DeepSkyBlue}{rgb}{0.000,0.750,1.000}

\tikzstyle{mystyle}=[scale= \PicSize,  
baseline=(current bounding box.center),
nodedescr/.style={fill=white,inner sep=2.5pt},
nodedescr2/.style={draw,circle,fill=white},
description/.style={fill=white,inner sep=2pt},
cancer loop/.style={preaction={draw=white, -,line width=6pt}, line width=1.1pt},
pot1 loop/.style={preaction={draw=white, -,line width=6pt}, red, line width=1.1pt},
pot2 loop/.style={preaction={draw=white, -,line width=6pt}, green, line width=1.1pt},
pot1/.style={preaction={draw=white, -,line width=6pt}, gray, solid, line width=1pt}, 
pot2/.style={preaction={draw=white, -,line width=6pt}, gray, solid, line width=1pt},
inPot/.style={ out=45,in=90,distance=3cm, min distance=2cm, line width=1.1pt, black!75},
inPot1/.style={ out=45,in=120,distance=3cm, min distance=2cm, line width=1.1pt, black!75},
inPot2/.style={ out=-45,in=-120,distance=3cm, min distance=2cm, line width=1.1pt, black!75},
in100Pot1/.style={ out=80,in=90,distance=3cm, min distance=2cm, line width=1.1pt, darkgreen},
in100Pot2/.style={ out=-80,in=-90,distance=3cm, min distance=2cm, line width=1.1pt, brown},
in2Pot1/.style={ out=45,in=90,distance=3cm, min distance=2cm, line width=1.1pt, black!75},
in2Pot2/.style={ out=-65,in=-90,distance=3cm, min distance=2cm, line width=1.1pt, black!75},
selfloop1/.style={ out=45,in=120,distance=6cm, min distance=4cm, line width=1.1pt, red},
selfloop2/.style={ out=-45,in=-120,distance=6cm, min distance=4cm, line width=1.1pt, blue},
loop1red/.style={ out=45,in=120,distance=6cm, min distance=4cm, line width=1.1pt, red},
loop2red/.style={ out=-45,in=-80,distance=6cm, min distance=4cm, line width=1.1pt, red},
cross line/.style={preaction={draw=white, -,	line width=6pt}},
jump1/.style={ out=45,in=135,distance=4cm, min distance=2cm, line width=1.1pt, red, dashed},
jump2/.style={ out=-45,in=-135,distance=4cm, min distance=2cm, line width=1.1pt, blue, dashed},
jumpover/.style={ out=80,in=90,distance=4cm, min distance=3cm, line width=1.1pt, black, dashed},
stochastic jump/.style={>=latex,>->,dashed, line width=1.1pt, green}
]

\def\PicSize{0.5} 

\def\nexttoPicSize2{6.0cm}

\def\oriPicSizeA{2.8cm}

\def\oriPicSizeB{3.1cm}

\numberwithin{equation}{section}

\usepackage[margin=1.4in]{geometry} 

\usepackage[parfill]{parskip}    

\begin{document}


\title{How to Grow an Organism Inside-Out: \\
Evolution of an internal skeleton from an external skeleton in bilateral organisms
}

\author{Eric Werner \thanks{\scriptsize Balliol Graduate Centre, Oxford Advanced Research Foundation (http://oarf.org), Cellnomica, Inc. (http://cellnomica.com). Thanks: Francis Hitching for careful editing and encouragement.  Martin Brasier for helpful discussions. We are grateful to Cellnomica, Inc. for use of its software to model multicellular development, both bilateral symmetry and inside-out development of organisms.   \copyright Werner 2012.  All rights reserved. }\\
University of Oxford\\
Department of Physiology, Anatomy and Genetics, \\
and Department of Computer Science, \\
Le Gros Clark Building, 
South Parks Road, 
Oxford OX1 3QX  \\
email:  eric.werner@dpag.ox.ac.uk\\
}

\date{ } 

\maketitle

\thispagestyle{empty}

\begin{center}
\textbf{Abstract}

\begin{quote}
\it 
An intriguing unanswered question about the evolution of bilateral animals with internal skeletons is how an internal skeleton evolved in the first place.  Computational modeling of the development of bilateral symmetric organisms suggests an answer to this question.  Our hypothesis is that an internal skeleton may have evolved from a bilaterally symmetric ancestor with an external skeleton.  By growing the organism inside-out an external skeleton becomes an internal skeleton. Our hypothesis is supported by a computational theory of bilateral symmetry that allows us to model and simulate this process. Inside-out development is achieved by an orientation switch.  Given the development of two bilateral founder cells that generate a bilateral organism, a mutation that reverses the internal mirror orientation of those bilateral founder cells leads to inside-out development.  The new orientation is epigenetically inherited by all progeny.  A key insight is that each cell contained in the newly evolved organism with the internal skeleton develops according to the very same downstream developmental control network that directs the development of its exoskeletal ancestor. The networks and their genomes are are identical, but the interpretation is different because of the cell's inverted orientation.  The result is inside-out bilateral symmetric development generating an inside-out organism with an internal skeleton.  
\end{quote}
\end{center}

\pagebreak

\pagenumbering{roman}
\setcounter{page}{1}
\tableofcontents

\pagenumbering{arabic}

\section{Introduction} 

An intriguing unanswered question about the evolution of bilateral animals with internal skeletons is how an internal skeleton evolved in the first place.  Computational modeling of the development of bilateral symmetric organisms suggests an answer to this question.  

Our hypothesis is that an internal skeleton may have evolved from a bilaterally symmetric ancestor with an external skeleton.  By growing the organism inside-out an external skeleton becomes an internal skeleton. Some early worms had an external skeleton.  Since a relatively simple epigenetic switch can make a bilateral organism grow inside-out,   applying this transformation to such primitive organisms with an external skeleton may have resulted in archaic organisms with an external skeleton having an internal skeleton. Our hypothesis is supported by a computational theory of bilateral symmetry that allows us to model and simulate this process. 

In particular, we hypothesize that a cellular, epigenetic transformation was responsible for the early evolution of the internal skeleton in metazoans. 
Inside-out development is achieved by an orientation switch.  Given the development of two bilateral founder cells that generate a bilateral organism, a mutation that reverses the internal mirror orientation of those bilateral founder cells leads to inside-out development.  The new orientation is epigenetically inherited by all progeny \cite{Werner2012a}.  

Inside-out growth utilizes developmental control networks that are interpreted by the cell's interpretive executive system (IES).  The IES interprets and executes the instructions in the genome \cite{Werner2011a, Werner2011b}.  

A key insight is that each cell contained in the newly evolved organism with the internal skeleton develops according to the very same downstream developmental control network that directs the development of its exoskeletal ancestor. The networks and their genomes are identical, but the interpretation is different because of the cell's inverted orientation.  The result is inside-out bilateral symmetric development generating an inside-out organism with an internal skeleton. 

First we will briefly explain how organisms develop bilaterally. Then we will explain how make a bilateral organism grow inside-out growth.  We model and simulate the multicellular development of a primitive organism with an external skeleton that is transformed into a organism with a central internal skeleton by means of inside-out growth.  

As a result we have a theory of how an internal skeleton may have evolved from an external skeleton.

\section{How bilateral development works}
In \cite{Werner2012a} we presented a theory of bilateral symmetric development of multicellular organisms. The core of that theory is based on the assumption that cells have an internal coordinate system that gives them an orientation in space.  When a cell makes an oriented division it results in two bilateral founder cells with mirror orientations.  The oriented division may happen directly or by means of a cell signaling protocol. The result is the same, namely, two daughter cells whose orientation axes mirror one another.  

\begin{figure}[H]
\begin{centering}
\hspace{0.5cm}
\subfloat[Part1][Founder cell's orientation]{\includegraphics[scale=0.33]{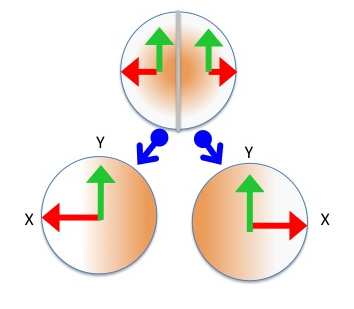}\label{fig:FounderCellsBB}}
\subfloat[Part1][A simple MCO]{\includegraphics[scale=0.33]{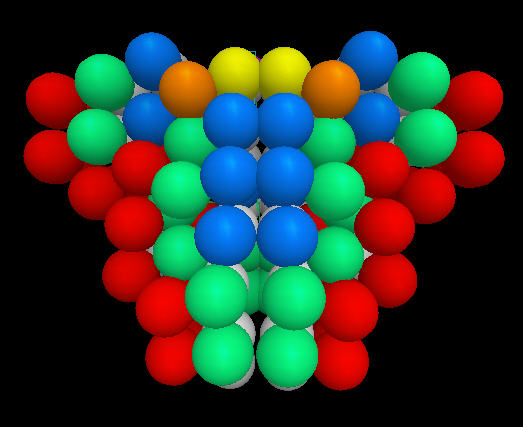}\label{fig:FemaleBase1}}
\hspace{0.5cm}
\subfloat[Part 1][Cell orientation view]{\includegraphics[scale=0.4]{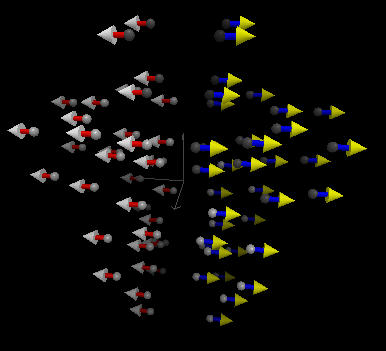} \label{fig:GynaderArrows}}
\end{centering}
\caption{
{\bf Oriented Back-to-Back cell division and the resulting bilateral organism.}  
\it \small Bilateral symmetric growth results from a mirror orientation of the daughter cells from a founder cell. The Back-to-Back orientation is epigenetically inherited during each division. Each directive in the organism's developmental control network that utilizes orientation along the X-axis is interpreted in the opposite direction in each generated polar opposite, mirror cell.  The result is a bilaterally symmetric multi-cellular organism. The first Fig.\ref{fig:FounderCellsBB} shows the establishment of two bilateral founder cells. Fig.\ref{fig:FemaleBase1} illustrates a possible resulting bilateral multicellular organism. The third Fig.\ref{fig:GynaderArrows} shows the internal epigenetically inherited opposite orientations of the cells in the two bilateral halves of the organism. (This figure is adapted from \cite{Werner2012a})}
\label{fig:BBSymmetryViews} 
\end{figure}

\subsubsection{The network perspective}
The oriented division produces two mirror daughter cells in identical developmental network control states realized in identical genomes. Even though each daughter founder cell is controlled by an equivalent developmental control network or cene \cite{Werner2011a}, the interpretation each cell gives to its developmental network is different. The interpretation depends on the cell's orientation state.   The alternative orientation-relative interpretations identical control networks, results in the development of bilaterally symmetric organism (see \cite{Werner2012a} for details).  

\subsubsection{The simulation perspective}
Cells are viewed as agents in a distributed multiagent system. Development is viewed as highly distributed, dynamic multiagent processes of communicating and cooperating social, physical agents acting in continuous space-time. The genome, the developmental control network, the cell differentiation state, the cell control  state, the cell physics, and the epigenetic, cell orientation state are all modeled. Each cell in the multicellular developing organism interprets and executes its developmental network in parallel with the other cells. The cells have the capacity to communicate and cooperate with one another via various signaling protocols and strategies. 

\section{Inside-out growth transforms an external skeleton to an internal skeleton}
\begin{figure}[H]
\subfloat[{\bf Regular}  ]{
\includegraphics[scale=.28]{externalBackbone-2rowOriBB3DRedTailcolor116.jpg}\label{fig:BBColoredMCO}}
\subfloat[{\bf Inside-out}  ]{
\includegraphics[scale=.28]{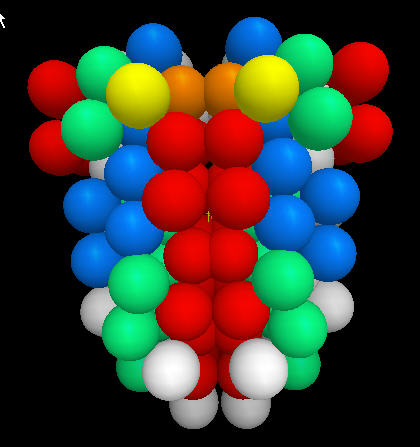}
\label{fig:FFColoredMCO}}
\subfloat[{\bf External skeleton}  ]{\includegraphics[scale=.28]{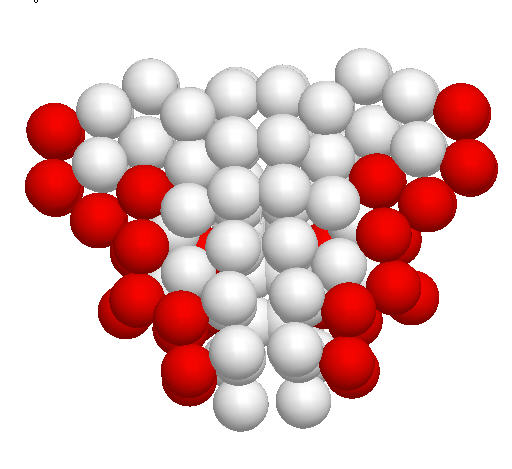}
\label{fig:BBRedWhiteMCO}}
\subfloat[{\bf Internal skeleton}  ]{\includegraphics[scale=.28]{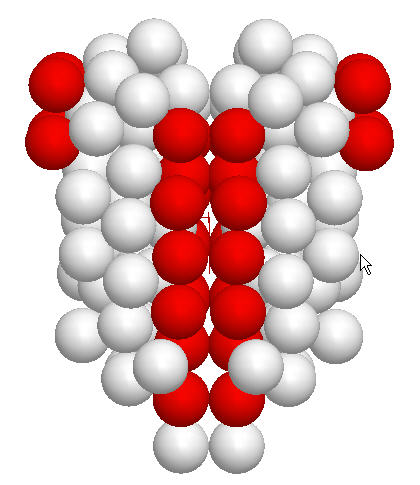}
\label{fig:FFRedWhiteMCO}}
\caption{
{\bf Inside-out growth transforms an external to an internal skeleton.} \it \small  Views of two pairs of multicellular organisms generated by a computational simulation.   On the left \ref{fig:BBColoredMCO} and \ref{fig:FFColoredMCO} are the cell differentiation views where cell colors represent different cell differentiation states. On the right the skeletal cells are emphasized in \ref{fig:BBRedWhiteMCO} and \ref{fig:FFRedWhiteMCO}.  A simple mutation results in an orientation switch in bilateral founder cells that then epigenetically generates the opposite, inside-out development in the organism.  }
\label{fig:BBFFColoredRedWhite}
\end{figure} 

A bilaterally symmetric organism develops bilaterally because at some point two founder cells have been generated such that their orientations mirror one another. Bilateral development follows because the cells on each side of the bilateral organism have mirror orientations.    When the initial mirror cell orientation is reversed by some means then the founder cells generate an organism that grows inside-out. It continues to grow inside-out so long as it remains viable.  

\autoref{fig:BBFFColoredRedWhite} illustrates the effect for a simple organism.  The first pair shows how the transformation looks in a normal view where the cell colors indicate different cell differentiation states. The second pair of organisms emphasizes the skeletal cells in red and the other non-skeletal cells are in white.  It shows how that inside-out growth can transform an organism with an external skeleton to an organism with an internal skeleton. 

\subsection{Inside-Out Growth Explained} 
When a mutation results in a reversal of the orientation of the bilateral founder cell, the effect is that the multicellular organism grows inside-out. Each directive in the developmental control network is now interpreted and executed in the opposite direction from before resulting in inside-out growth of the bilateral organism.

\begin{figure}[H]
\begin{centering}
\subfloat[Part 1][\bf External skeleton]{\includegraphics[scale=.55]{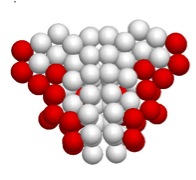}\label{fig:RedOutside}}
\subfloat[Part 2][\bf Orientation switch]{\includegraphics[scale=.50]{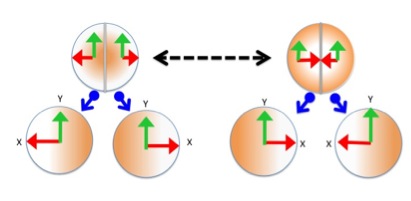}\label{fig:BB2FFtransform}}
\subfloat[Part 4][\bf Internal skeleton]{\includegraphics[scale=.60]{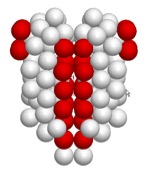}\label{fig:RedInside}}
\end{centering}
\caption{
{\bf Orientation transform results in inside-out growth.}  
\it \small The bilateral multicellular organism on the left Fig.\ref{fig:RedOutside}, where the red cells are on the outside, develops from the founder cells next to its right in Fig.\ref{fig:BB2FFtransform} that have a Back-to-Back orientation.   The reversal from a Back-to-Back orientation  to a  Face-to-Face orientation in the founder cells Fig.\ref{fig:BB2FFtransform} is epigenetically inherited in the progeny.  The result of this orientation switch is inside-out development seen in the organism on the right Fig.\ref{fig:RedInside} where the red cells are on the inside. The transformation is reversible going in either direction depending only on the initial orientation of the bilateral founder cells. (This figure is adapted from \cite{Werner2012a})}
\label{fig:InsideOutOri}
\end{figure}

In \autoref{fig:InsideOutOri} above, a reversal of the X-axis (in red) from a Back-to-Back (on the left) to a Face-to-Face orientation (on the right) in the founder cell results in an opposing Face-to-Face orientation in the daughter cells. This new orientation is epigenetically inherited in the progeny. The result is inside-out development. 

\section{A new evolutionary space}
The developmental process by which bilateral symmetry and symmetry inversions are established is fundamentally different from simple  traditional genetically based processes.  A standard genetic mutation changes a protein that is pleiotropic and is rarely associated with a particular phenotype.  Here a genetic mutation may induce an orientation inversion and thereby generate a whole new distributed developmental process that effects all the progeny of the transformed cell.  The transformation is inherited by the daughter cells at each division.  

Genes must be activated prior to each division so as to manufacture the self-organizing components needed to maintain the internal coordinate system of the cell and thereby recreate the orientation of the parent cell in its daughter cells. However, the gene products not only self-organize but interact with the given structure of the cell using that structure as a template to maintain the orientation and coordinate system in the new daughter cells.  Hence, the inheritance of the orientation structure of the cell is largely epigenetic while at the same time utilizing the parts produced by the genome.  

Bilateral symmetry is achieved by a relatively simple mechanism \cite{Werner2012a}.  This may explain the rapid evolution of bilaterally symmetric organisms, as well as internal skeletal life forms in the pre-Cambrian and Cambrian.  The process is reversible so it is also possible that exoskeletons resulted from inverted inside-out development of bilateral organisms with an internal skeleton. 

\subsection{Evolutionary potential of sub-symmetry inversion}
Sub-symmetries can also be generated by oriented divisions that are downstream and in the same dimension as the first oriented division.

If we have an addition reversal of orientation in the next generation, then only those offsprings downstream in the control network will invert their development.  This is best seen by an example. 
\begin{figure}[H]
\centering
\subfloat[Part 9][RGB-ff-bb]{\includegraphics[height=\oriPicSizeB]{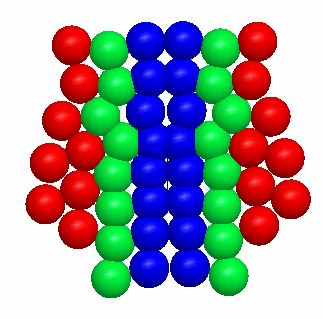}\label{fig:RGB-ff-bb}}
\hspace{0.3cm}
\subfloat[Part 11][RGBa-ff-bb]{\includegraphics[height=\oriPicSizeB]{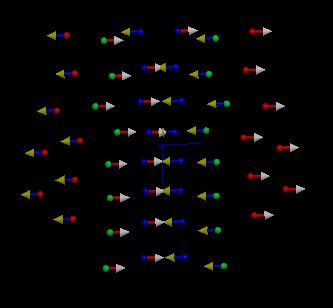}\label{fig:RGBa-ff-bb}}
\hspace{0.5cm}
\subfloat[Part 13][RBGa-ff-ff]{\includegraphics[height=\oriPicSizeB]{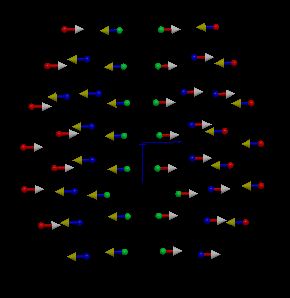}\label{fig:RBGa-ff-ff}} 
\subfloat[Part 15][RBG-ff-ff]{\includegraphics[height=\oriPicSizeB]{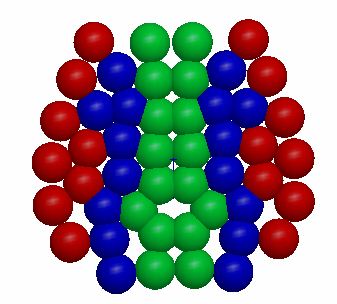}\label{fig:RBG-ff-ff}}\\ 
\subfloat[Part 8][BGR-bb-bb]{\includegraphics[height=\oriPicSizeA]{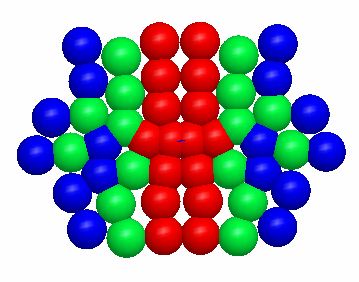}\label{fig:BGR-bb-bb}}  
\subfloat[Part 6][BGRa-bb-bb]{\includegraphics[height=\oriPicSizeA]{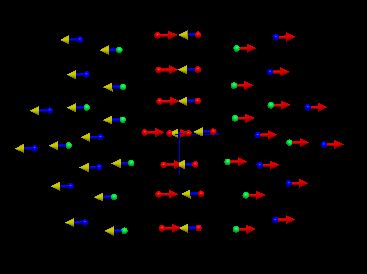}\label{fig:BGRa-bb-bb}} 
\hspace{0.1cm}
\subfloat[Part 3][GBRa-bb-ff]{\includegraphics[height=\oriPicSizeA]{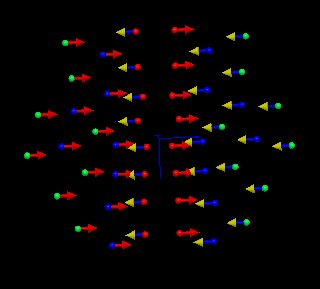}\label{fig:GBRa-bb-ff}}
\subfloat[Part 1][GBR-bb-ff]{\includegraphics[height=\oriPicSizeA]{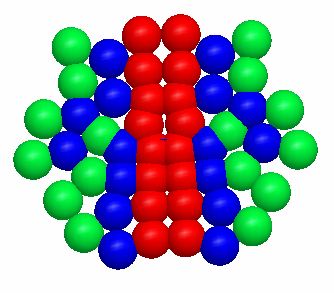}\label{fig:GBR-bb-ff}}
\caption{{\bf Transformations of symmetries and sub-symmtries.} \it \small These multicellular structures illustrate different combinations of orientation switches leading to different combinations of layers of bilaterally symmetric tissue.  Even these simple stylized examples show the significant evolutionary potential of variations generated by transformations of cell orientation. The arrows show the different states of orientation of the adjacent cell tissue. {\bf Notation:} B = Blue, G = Green, R = Red, bb = back-to-back, ff = face-to-face.  Hence, GBR-bb-ff means that a Green Blue Red pattern of cells results from a back-to-back followed by a face-to-face cell division.  }
\label{fig:SubSymmetries}
\end{figure}

Imagine an organism with three layers of major cell types indicated by Red, Green, Blue.  For example in Fig.\ref{fig:RGB-ff-bb}, RED is on the outside edge and BLUE is in the center as follows: \textcolor{red}{RED}-\textcolor{green}{GREEN}-\textcolor{blue}{BLUE} || \textcolor{blue}{BLUE}-\textcolor{green}{GREEN}-\textcolor{red}{RED} where || is the plane of symmetry.  If we do a reversal in the upper most oriented division, the the order is reversed:  \textcolor{blue}{BLUE}-\textcolor{green}{GREEN}-\textcolor{red}{RED} || \textcolor{red}{RED}-\textcolor{green}{GREEN}-\textcolor{blue}{BLUE} as in Fig.\ref{fig:BGR-bb-bb}.  Now RED is in the center and BLUE is on the outside edge.  

If, however, we reverse the order of symmetry in a second oriented division then only the Green and Blue grow inside-out. So starting with the same state as before in Fig.\ref{fig:RGB-ff-bb} where BLUE is in the center as follows: \textcolor{red}{RED}-\textcolor{green}{GREEN}-\textcolor{blue}{BLUE} || \textcolor{blue}{BLUE}-\textcolor{green}{GREEN}-\textcolor{red}{RED} a secondary orientation switch only reverses the order of the middle layers, exchanging GREEN with BLUE to give \textcolor{red}{RED}-\textcolor{blue}{BLUE}-\textcolor{green}{GREEN} || \textcolor{green}{GREEN}-\textcolor{blue}{BLUE}-\textcolor{red}{RED} as in Fig.\ref{fig:RBG-ff-ff}. Note that RED stays untransformed on the outside edge while only the center is transformed. 

If instead we start with the state Fig.\ref{fig:RBG-ff-ff} which is  \textcolor{red}{RED}-\textcolor{blue}{BLUE}-\textcolor{green}{GREEN} || \textcolor{green}{GREEN}-\textcolor{blue}{BLUE}-\textcolor{red}{RED} and switch the first major orientated division we then get GREEN on the outside to give Fig.\ref{fig:GBR-bb-ff} which is \textcolor{green}{GREEN}-\textcolor{blue}{BLUE}-\textcolor{red}{RED} || \textcolor{red}{RED}-\textcolor{blue}{BLUE}-\textcolor{green}{GREEN}.  In other words, the inside-out growth switch can be applied at any level where there is oriented division. But it only applies to cell states in network paths that are downstream of the orientation switch.  Hence, we can get every possible combination of order of tissue layers depending on the orientation switches.   

Therefore, orientation switching has implications for ordering of tissue layers within organisms. Mutations that result in orientation switches can reverse the location of internal tissue in bilateral organisms without the whole organism growing inside-out.  This open up a new route for the evolution of multicellular organisms. 

\section{Conclusion}

We shown that it is possible that the internal skeleton of multicellular organisms may have evolved from an ancestor with an external skeleton.  Computational modeling and simulation show that this is possible by way of a mutation that leads to an orientation switch in the bilateral founder cells. This change in cell orientation is inherited by the progeny resulting to the previously exoskeletal organism growing inside-out transforming the formally external skeleton into an internal skeleton.  

This major evolutionary step is a combination of the interaction of genetics, epigenetics, bilateral symmetry and developmental control networks.   A gene mutation may be responsible for the initial orientation switch.   The inheritance of the orientation in daughter cells from their parent cell is epigenetic but is supported by genes that prior to each cell division produce the parts that make up the molecular orientation structures necessary for oriented division. The control of bilateral development in each bilateral body half is directed by developmental control networks.  

Evolution by cell orientation transformation is a fundamentally different process from traditional evolution by genetic mutation.  It is also different from evolution by developmental network mutation \cite{Werner2011a, Werner2011b}. The creation and transformation of symmetries and sub-symmetries opens up alternative evolutionary pathways.  It offers a vast new potential evolutionary space of possible morphologies and functions for evolving multicellular organisms.  See \cite{Werner2012a} for some examples.  This evolutionary space predicted by our theory of symmetry and symmetry transformation has been not explored up until now.  Like a newly discovered country waiting to be explored, it leads to open questions and new areas of possible research, .   

\section{Materials and methods}
\label{Methods}
Cells are viewed as agents in a distributed multiagent system. Development is viewed as highly distributed, dynamic multiagent processes of communicating and cooperating social, physical agents acting in continuous space-time. The genome, the developmental control network, the cell differentiation state, the cell control  state, the cell physics, and the epigenetic, cell orientation state are all modeled. Each cell in the multicellular developing organism interprets and executes its developmental network in parallel with the other cells. The cells have the capacity to communicate and cooperate with one another via various signaling protocols and strategies. 

We used Cellnomica's Software Suite (http://cellnomica.com) to model and simulate multicellular development in space-time.  The developmental control networks, cell orientation, bilateral symmetry, sub-symmetries, symmetry breaking and inside-out mutations were all tested using Cellnomica's Software Suite. Each of the concepts discussed was  modeled and simulated with artificial genomes that generated multicellular bilaterally symmetric organisms starting from a single cell. Mutations to the developmental control networks that resulted in the reversal of the axis of orientation leading to inside out growth and sub-symmetries were also modeled. The illustrations of multi-cellular systems are screenshots of cells that developed in virtual 4-dimensional space-time using the Cellnomica's software.

\addcontentsline{toc}{section}{References}

\footnotesize 
\bibliographystyle{abbrv}
\bibliography{BilateralSymArXiv}

 Let me know if your article is especially relevant and should be included.  
\end{document}